\documentclass[sigconf]{acmart}
\usepackage[ruled,vlined]{algorithm2e}
\usepackage{multirow}
\usepackage{dsfont}

\AtBeginDocument{%
  \providecommand\BibTeX{{%
    \normalfont B\kern-0.5em{\scshape i\kern-0.25em b}\kern-0.8em\TeX}}}

\setcopyright{acmcopyright}
\copyrightyear{2023}
\acmYear{2023}
\acmDOI{10.1145/1122445.1122456}

\acmConference[Singapore '23]{Singapore '23: ACM Symposium on Neural
  Gaze Detection}{February 27--March 3, 2023}{Singapore}
\acmBooktitle{Singapore '23: ACM Symposium on Neural Gaze Detection,
  February 27--March 3, 2023, Singapore}
\acmPrice{15.00}
\acmISBN{978-1-4503-XXXX-X/23/02}

\usepackage{subcaption}
\usepackage{array}
\usepackage{multirow}
\usepackage{todonotes}
\usepackage{enumitem}

\renewcommand{\todo}[1]{\iffalse #1 \fi[TODO]}

\begin{document}

\title{Personalizing Intervened Network for Long-tailed Sequential User Behavior Modeling}

\author{Zheqi Lv$^{*\dagger}$, Feng Wang$^{*\ddagger}$, Shengyu Zhang$^{*\dagger}$, Kun Kuang$^\dagger$, Hongxia Yang$^\ddagger$, Fei Wu$^\dagger$}
\thanks{*These authors contributed equally to this research.}
\affiliation{
  \institution{
\textsuperscript{$\dagger$}Zhejiang University \city{Hang Zhou}\country{China}
\textsuperscript{$\ddagger$}DAMO Academy, Alibaba Group \city{Hang Zhou}\country{China}
}}

\newcommand{\vpara}[1]{\vspace{0.05in}\noindent\textbf{#1 }}
\renewcommand{\shortauthors}{Trovato and Tobin, et al.}




\begin{abstract}
    In an era of information explosion, recommendation systems play an important role in people's daily life by facilitating content exploration. It is known that user activeness, \textit{i.e.}, number of behaviors, tends to follow a long-tail distribution, where the majority of users are with low activeness. In practice, we observe that tail users suffer from significantly lower-quality recommendation than the head users after joint training. We further identify that a model trained on tail users separately still achieve inferior results due to limited data. Though long-tail distributions are ubiquitous in recommendation systems, improving the recommendation performance on the tail users still remains challenge in both research and industry. Directly applying related methods on long-tail distribution might be at risk of hurting the experience of head users, which is less affordable since a small portion of head users with high activeness contribute a considerate portion of platform revenue. In this paper, we propose a novel approach that significantly improves the recommendation performance of the tail users while achieving at least comparable performance for the head users over the base model. The essence of this approach is a novel Gradient Aggregation technique that learns common knowledge shared by all users into a backbone model, followed by separate plugin prediction networks for the head users and the tail users personalization. As for common knowledge learning, we leverage the backward adjustment from the causality theory for deconfounding the gradient estimation and thus shielding off the backbone training from the confounder, \textit{i.e.}, user activeness. We conduct extensive experiments on two public recommendation benchmark datasets and a large-scale industrial datasets collected from the Alipay platform. Empirical studies validate the rationality and effectiveness of our approach.
\end{abstract}

\maketitle

\section{Introduction}

\begin{figure}[h]
    \centering
    \begin{subfigure}[b]{0.234\textwidth}
           \centering
           \includegraphics[width=\textwidth]{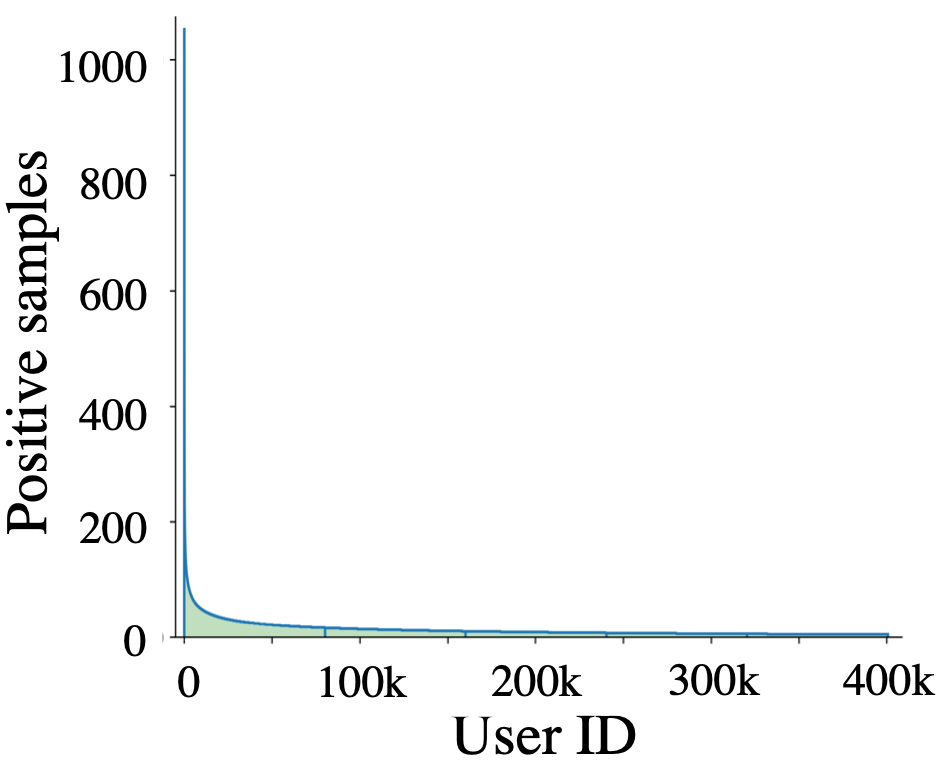}
           %\vspace{-2mm}
            \caption{}
            \label{fig:a}
    \end{subfigure}
    \begin{subfigure}[b]{0.234\textwidth}
            \centering
            \includegraphics[width=\textwidth]{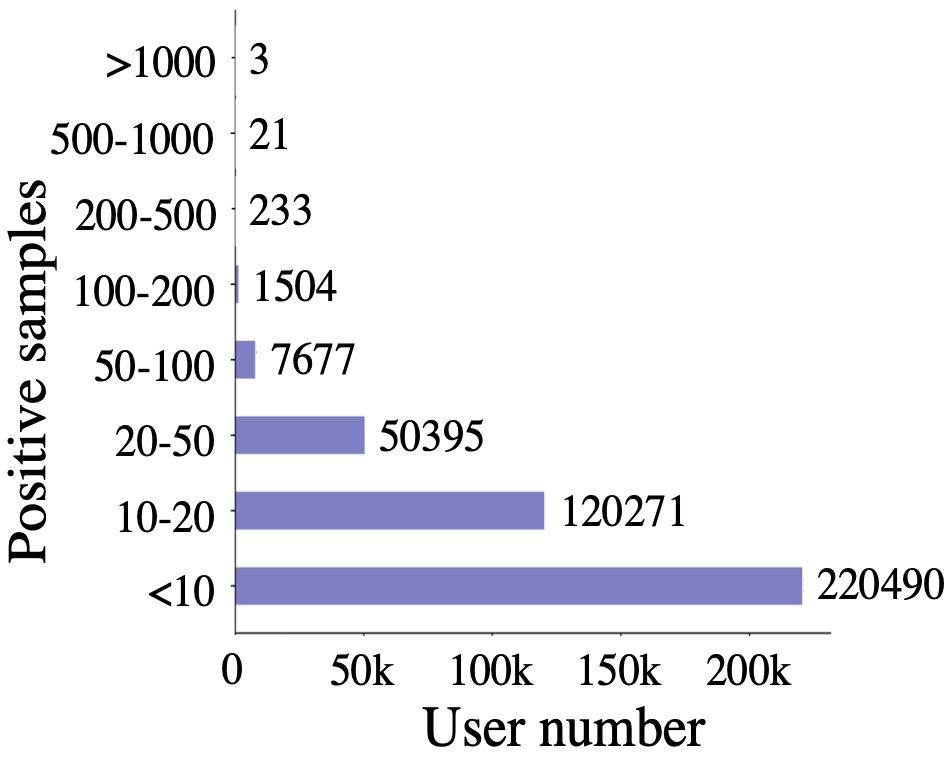}
            %\vspace{-2mm}
            \caption{}
            \label{fig:b}
    \end{subfigure}
    %\vspace{-2mm}
    \caption{The long-tail effect on Alipay dataset.}
    \label{fig:longTailEffect}
%\vspace{-2mm}
\end{figure}

With the rapid development of the Internet in recent years, recommendation systems play a vital role in people's daily life. In order to improve the accuracy of the recommendation system, researchers have designed varying recommendation algorithms ranging from collaborative filtering and matrix factorization-based FM\cite{FM} to Wide\&Deep\cite{wide&deep}, DeepFM\cite{deepfm} and other deep learning methods. Since users' behaviors naturally show chronological dependencies, many sequencial recommendation algorithms have been proposed, including, but not limited to, GRU4Rec\cite{GRU4Rec}, DIN\cite{DIN}, DIEN, SASRec\cite{SASRec}, etc. 
One of the most enticing merits of sequential models is that we can dynamically update the recommendation list as long as users interact with new items.
% These algorithms have allowed the recommendation system to develop rapidly in just a few years. 
Despite their great advances, we identify that existing methods still achieve less satisfied recommendation quality for non-active users. In industrial recommendation systems, it is known that user activeness tends to form a long-tail distribution. For example, as shown in Fig. \ref{fig:longTailEffect}, the majority of users have significantly less number of behaviors than head users. Apparently, to improve recommendation quality of most users, 

it is of paramount importance to pay attention to the tail users with low activeness.

\begin{figure*}[h]
    \centering
    \includegraphics[width=0.65\textwidth]{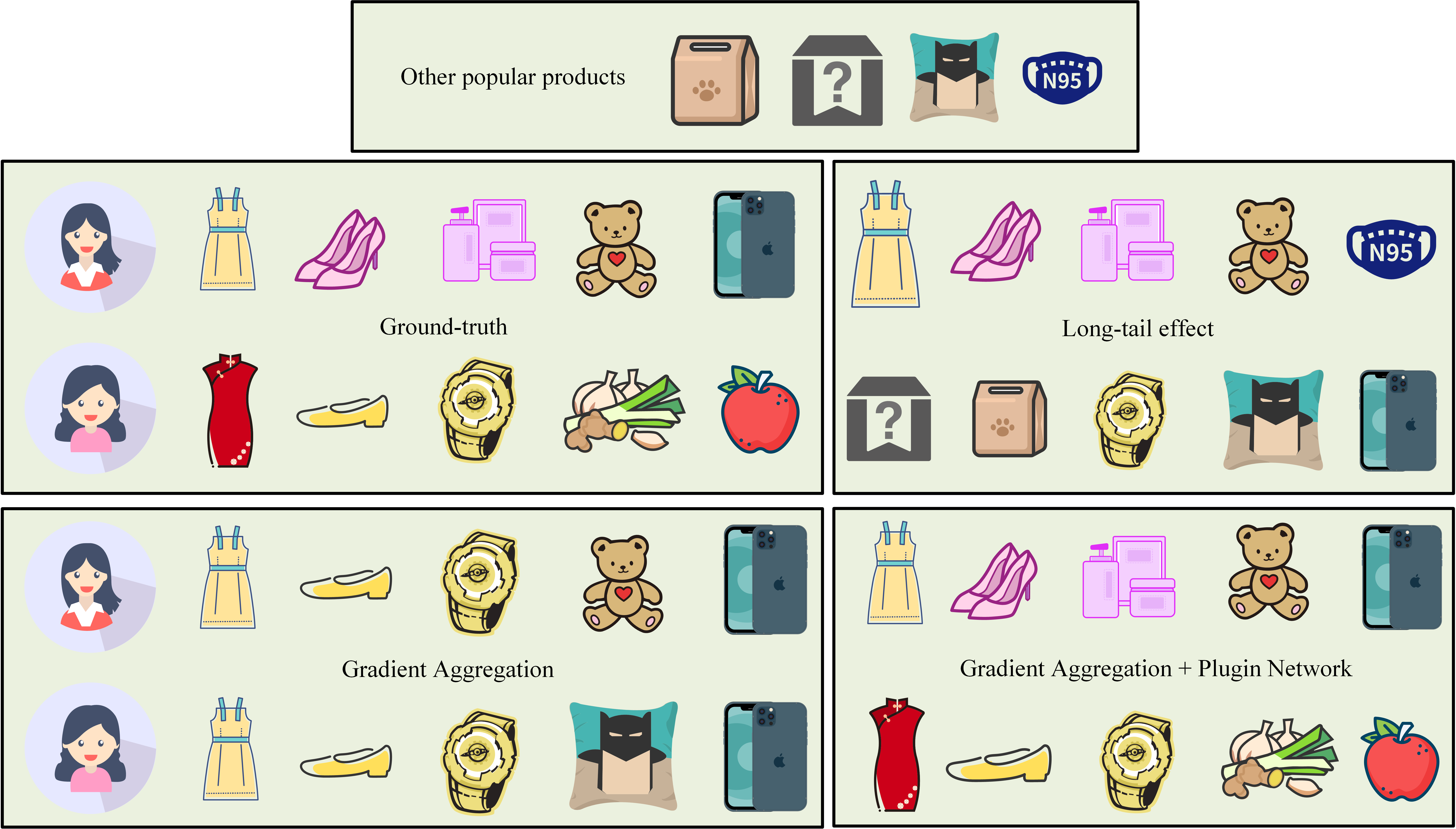}
    %\vspace{-2mm}
    % \label{fig_example}
    \caption{A toy example of the long-tail effect and the improvement brought by our method.
    In this case, there are two groups of women, one is young women with high activity and the other is middle-aged women with low activity. They have common preferences such as clothes and shoes but also some different preferences. Due to the long-tail effect, the preferences of low-activity women are difficult to capture, and the model will recommend some popular products for them.
    To address this problem, we extract generalization information via the gradient aggregation module so that the model can recommend common preferences such as clothes and shoes to low-activity women, although sometimes not her favorite style. The model's recommendation for the high-activity women and the low-activity women are more similar, so the performance of the model on high-activity women has decreased.
    Next, we train a plugin network for each group of women. The plugin network captures the group-specific personalization information such as different styles of clothes and shoes and other unpopular preference.}
\label{fig:toy}
%\vspace{-2mm}
\end{figure*}
Investigating long-tail distributions has gained much attention in the literature~\cite{long-tail_survey}, especially in computer vision~\cite{rebalance1,transfer1,moduleImpr1}.

However, these methods cannot be directly applied to the recommendation domain due to its unique challenges. Specifically,
in the recommendation system, the behaviors of tail users are quite limited, leading to less adequate preference understanding. Moreover, we identify that models are easily biased towards head users during joint training and thus make the problem even more severe.
As for separate training for tail users, we conduct an empirical study and find that the separately trained model still achieves inferior results due to the limited amount of data. 
At present, studying tail users in recommendation systems is still a nascent research field, such as Grad-Transfer\cite{gradtrans}. 
Despite the performance improvement on tail users, most of them are at risk of sacrificing the performance of head users. Note that hurting the experiences of head users is less affordable in industrial environments due to the large amount of platform revenue contributed by active users. 
Another line of related works is cold-start recommendation~\cite{MeLU,MetaEmb,MAMO}, which aims to improve the performance of users that are not observed during training. We differ from these works fundamentally by improving the performance of observed tail users during joint training.

We argue that improving the model's performance on the tail user base does not necessarily degrade the model's performance on the short-head user base. Similar users (no matter how active they are) should all benefit from joint training through collaborative filtering. 
Besides these common knowledge shared by similar users, each user has their personalized preferences.
Therefore, we argue that one possible solution is to extracting common knowledge that is shared by all users and is independent from user activeness while retaining the personalized information for different users.
Upon the above analysis, we devise a gradient aggregation framework for long-tailed sequential user behavior modeling. To shield off the common knowledge learning from user activeness, we borrow back-door adjustment~\cite{neuberg2003causality} technique from the causality theory. In particular, we view the user activeness as a confounder and deconfound the model training via causal intervention. To instantiate back-door adjustment, a grouping strategy and a gradient aggregation strategy are proposed. We group each group of data according to the activity divided by the time window, and at the same time ensure that the amount of data in each group is as equal as possible. During training, we intervene the gradient estimation via back-door adjustment, leading to a activeness-independent training. Upon the causally learned trunk that encodes the common knowledge, we devise multiple plugin networks for group-specific personalization.
Through the architecture of the trunk network and the plugin network, the model not only has strong general reasoning ability, but also has good individual reasoning ability. In Fig. \ref{fig:toy}, we give a toy example of long-tail effect of a real case in Alipay platform and show the improvement brought by our propose method. Moreover, we conduct extensive experiments on industrial datasets to demonstrate the effectiveness of our method. 
In this paper, our main contributions are as follows:
% \vspace{-6mm}
\begin{itemize}
% \vspace{-6mm}
    \item We investigate how to improve the recommendation quality of tail non-active users without sacrificing the performance on active users. We identify disentangle the common knowledge learning and personalization is a plausible solution to the problem.
    \item We learn a recommendation trunk that encodes the common knowledge via back-door adjustment borrowed from causality theory. With causal intervention, the estimated gradient should be independent from user activeness. Upon the causal trunk, we devise several plugin networks for group-specific personalization.
    \item We completed experiments on Movielens, Amazon, and Alipay datasets. Experiments show that our method is practical and effective.
\end{itemize}

\section{Related work}

\subsection{Long-tail}
Deep long-tail learning is one of the most challenging problems in deep learning, and its goal is to train well-performing deep learning models from data that follow a long-tailed class distribution. Long-tail class imbalance tends to limit the usefulness of deep learning-based models in real-world applications, as they tend to be biased towards dominant classes and perform poorly on tail classes.\cite{long-tail_survey} In order to solve this problem, researchers have carried out a lot of research in recent years, and have made gratifying progress in the field of deep long-tail learning. Existing research on deep long-tail learning is mainly divided into three categories, namely class rebalancing\cite{rebalance1,rebalance2,rebalance3,rebalance4}, information augmentation\cite{transfer1,transfer2,transfer3,transfer4,gradtrans}, and module improvement\cite{moduleImpr1,moduleImpr2,moduleImpr3,moduleImpr4}.

The recommendation system has made great progress in the past few years, but in real recommendation scenarios, the datasets often satisfy the long-tail distribution, which also makes the current recommendation algorithms always face the long-tail problem. Grad-transfer\cite{gradtrans} adjusts the weight of each user during training through resample and gradient alignment, and adopts an adversarial learning method to avoid the model from using the sensitive information of user activity group in prediction. In this way, the performance of the model in the tail user group can be improved. Related studies also use meta-learning to improve the effect of cold start, such as MeLU\cite{MeLU}, MetaEmb\cite{MetaEmb}, MAMO\cite{MAMO}, MetaHIN\cite{MetaHIN}, PAML\cite{PAML}, GME\cite{GME} and MWUF\cite{MWUF}, explore initializing model parameters or embeddings with user and item side information. There are also some works on tail items, such as using causal\cite{causal} and a bipartite graph\cite{bipartite_graph}, tripartite graph\cite{tripartite_graph}, and extended tri-partite graph\cite{extended_tripartite_graph} to eliminate the popularity bias, and apply multi-objective optimization methods\cite{multi-task1,multi-task2,multi-task3,multi-task4} to recommender systems in order to improve the probability that the tail item is recommended.

Similarly, our research also focuses on how to improve the recommendation performance of the recommendation model for tail users. However, our method improves the model's performance on tail users without degrading its performance on head users.

\subsection{Gradient Surgery}
Gradient surgery is usually used in fields such as multi-task learning and domain generalization. It can use the knowledge learned on one task during the training process to improve the generalization performance of the model on other tasks. To achieve this, GradNorm\cite{GradNorm} dynamically scales the gradients to ensure that the gradients produced by each task contribute similarly to model training, thereby ensuring that the model trains multiple tasks at an approximate rate. PCGrad\cite{PCGrad} discusses the situation where the gradients of multiple tasks have conflicting components when the model is multi-task learning. It achieves soft fusion of two gradients with conflicting components by projecting between gradients of different tasks. GradVac\cite{GradVac} sets a separate gradient similarity objective for each task pair, making better use of inter-task correlations through adaptive gradient similarity. Agr-Rand \cite{Agr-Rand} solves domain generalization through gradient surgery. It determines how gradients are updated by comparing the sign bits between gradients across domains, resulting in a classifier that generalizes well across multiple domains.

Compared with the existing algorithms, we use causality theory to reduce the influence of user activity on the gradient, and design a plugin network to make up for the shortcomings of gradient surgery.

\section{Method}

\begin{figure*}[t]
  \centering
  \includegraphics[width=0.68\linewidth]{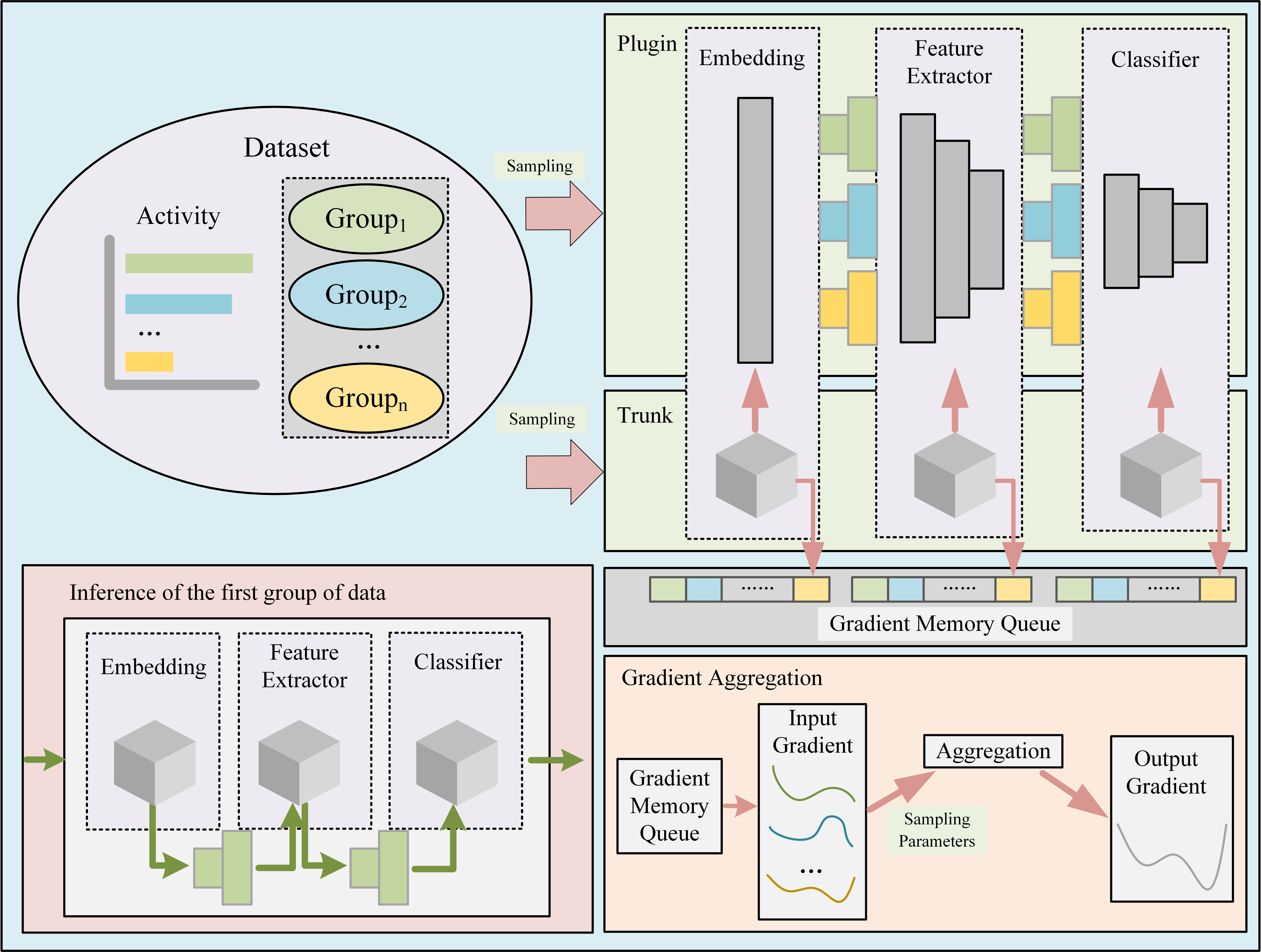}
  \vspace{-1.5mm}
  \caption{Architecture of the proposed method. (1) During training, our method consists of two stages. The first stage is to use all of the data for training and aggregate the gradients of each group to obtain a trunk network that can extract generalization information. The second stage is to insert the plugin network into the trunk network. The plugin network for each group is trained using only the data of this group. At the same time, the trunk network is still undergoing updates similar to the first stage. In general, the trunk network extracts the common information of all groups, and the plugin network extracts the personalized information of each group. (2) During inference, the plugin network is plugged to the trunk network to realize the fusion of general information inference and individual information inference.}
  \label{fig:arch}
\vspace{-1.5mm}
\end{figure*}
% \vspace{-1.5mm}
In this paper, we focus on the problem of long-tail distribution in recommendation in real industrial applications. To address this problem, we propose a novel framework to disentangle the common knowledge learning and personalization for improving the recommendation quality of the tail non-active users without sacrificing the performance on the head active users. Fig.~\ref{fig:arch} shows the architecture of our proposed method. Next, we will detail each component of the proposed method.

\subsection{Preliminaries}
% pcgrad
% Agr-Rand
The sequence recommendation model can generally be expressed as: supposing we have a user set $\boldsymbol{U}=\{u_1, u_2, ..., u_n\}$, and an item set $\boldsymbol{I}=\{i_1, i_2, ..., i_m\}$. The click sequence $\boldsymbol{S}=\{s^{u_1}, s^{u_2}, ..., s^{u_n}\}$ represents the $L$ items that the user $u$ has clicked recently so far. We randomly select some user-item pairs and the corresponding user behavior sequence $S^u$ from the data set and put them into the model to get a prediction score $\hat{y}_{u,i}$ between 0 and 1, that is,
\begin{align}
\label{eq:original_forward}
\hat{y}_{u,i}=f(u,i,s^u|\theta)
\end{align}
$f(\cdot)$ denotes the forward propagation function and $\theta$ represents the parameters of the model. The ground-truth $y_{u,i}\in\{0, 1\}$ is a binary variable, so we can use the binary cross entropy loss as the loss function which is commonly used in binary classification problems to calculate the loss. We use $\mathcal{D}$ to represent the entire dataset, then we can get the loss function $\mathcal{L}$ by
\begin{align}
\label{eq:loss_normal}
\mathcal{L}=-\frac{1}{|\mathcal{D}|}\sum_{i=1}^{|\mathcal{D}|}
y_i\log\hat{y}_{u,i}+(1-y_i)\log(1-\hat{y}_{u,i}) 
\end{align}
Then, we backpropagate the loss and update $f$, repeat several times to obtain the optimal parameter $\theta$, and finally a sequential recommendation model can be obtained.
\begin{align}
\label{eq:backward_normal}
\theta=\arg\min\limits_{\theta}\mathcal{L}
\end{align}

\subsection{Gradient Aggregation}
In the recommendation system, although the difference in user activity also affects the model's judgment of their preferences, users with different activity levels also have a lot of common information that can be extracted by the model. These common information can not only help the model perform better on the tail user group, but also help the model to better learn the preferences of high-active users. Since the common information is implicit and highly entangled with personalized information, it is hard to identify and extract them directly. To bridge the gap, we borrow the back-door adjustment technique \cite{neuberg2003causality} from the causal literature and deconfound the model training. Deconfounding means shielding the model training/prediction from possible confounders. Since the major focus is user activeness, i.e., the number of users' behaviors, we regard it as the confounder. Formally, by the Bayes rule, the original training gradients can be obtained by:
\begin{align}
	P(g \mid u,i, y_{u,i}) &= \sum_{z} P(g, z \mid u,i, y_{u,i}) \\
	&= \sum_{z} P(g \mid u, i, y_{u,i}, z) \underline{P(z \mid u,i, y_{u,i})},
\end{align}
where $g$ denotes the model gradients. $z$ denotes the confounder which is user activeness here. $P(\cdot)$ denotes the forward function that takes $u,i$ as the input and a deterministic backward function that estimates the gradient based on the prediction of the model and the ground-truth $y_{u,i}$. In the language of causality, the gradient estimation is confounded by the user activeness, leading to biased training that is favor of active users. Backward adjustment deconfounds the estimation by blocking the direct effect $z \rightarrow u,i,y_{u,i}$:
\begin{align}
	P(g \mid do(u,i, y_{u,i})) = \sum_{z} P(g | u,i, y_{u,i}, z) \underline{P(z)} \label{eq:backdoor}
\end{align}
where the proof can be found in~\cite{Pearl_2009,pearl1995causal,glymour2016causal}. In practice, the essence of the above adjusted equation is to estimate the gradient $P(g | u,i, y_{u,i}, z)$ for each $z$ value, and sum all gradients from different $z$ values with the prior probabilities.

To estimate the gradients per $z$ value, we divide users into $n$ groups according to the level of activeness. During training, each batch can be divided into $n$ mini-batches. When the $j$-th mini-batch in the $b$-th batch is trained with the data from the $j$-th group, we can get the gradient $\boldsymbol{g}^{|k|}_{i,j}$ produced by this mini-batch at the $k$-th layer of the model. Similarly, when we train all mini-batches in this batch, we can get a gradient set $\boldsymbol{G}^{|k|}_{i} = \{\boldsymbol{g}^{|k|}_{i,2}, \boldsymbol{g}^{|k|}_{i,2}, …, \boldsymbol{g}^{|k|}_{i,n}\}$. When a mini-batch in a batch completes backpropagation, we do not directly update the gradient generated by the mini-batch but put the gradient into a gradient memory queue $\boldsymbol{Q}^{|k|}$. After all mini-batches in this batch have completed backpropagation, we aggregate these gradients using Eq.~\ref{eq:backdoor}. The prior distribution of different $z$ values is empirically set to a uniform distribution. This is because we divide the users into equally-numbered groups and different user groups are treated as equally important so as to deal with the long-tail problems.
Formally, we obtain the deconfounded gradients as follows:

\begin{align}
\label{eq:gradient_aggregation}
\boldsymbol{g}^{|k|}_b=\frac{1}{n}\sum_{j=1}^n \frac{N}{n_j}\boldsymbol{g}^{|k|}_{b,j}.
\end{align}
Among them, $N$ represents the total number of samples in the dataset $\mathcal{D}$, $n_j$ represents the number of samples in the $j$-th user group. We use $f_t(\cdot)$ denotes the forward propagation function.

\subsection{Plugin Network}
After gradient aggregation, we can obtain a trunk network containing the common knowledge of all user groups. However, the group-specific knowledge is less modeled, which may cause degraded performance. To alleviate the problem, we introduce the plugin network for each group to learn the personalized knowledge from the data of this group.

\subsubsection{Naive Plugin Network}
% \\
\
\newline
A naive method is training a plugin network of the same size as the trunk network for each user group, with its model parameters denoted $\Delta\boldsymbol{\theta}$. When predicting users belonging to the $j$-th group, it is necessary to add the plugin network parameters $\Delta\theta_j$ and the trunk network parameters $\theta$ dedicated to the $j$-th group. The forward propagation function is, 
\begin{align}
\label{eq:naive_forward}
\hat{y}_{u,i}=f(u,i,j,s^u|\theta+\Delta\theta)
\end{align}

Furthermore, we split the model into embedding layers, feature extractors and classifiers. We use $E$ to denote the parameters of the embedding layer in the trunk network. Since the parameters of the embedding layer are very sparse, only using the data of a certain user group for training will lead to inaccurate mapping of the data in the feature space, we remove the embedding layer from the plugin network. Therefore, we only update the parameters of the feature extractor $\Delta\Phi_j$ and the classifier $\Delta W_j$ when we train the plugin network for the $j$-th user group. Formally, we predict the user-item pair of the $j$-th user group as follows, 
\begin{align}
\label{eq:naive_forward_detailed}
\hat{y}_{u,i}=f(u,i,j,s^u|E,\Phi+\Delta\Phi_j,W+\Delta W_j)
\end{align}
We compute the loss function as follows, where $\mathcal{L}_j$ denotes the data of the $j$-th group of users, 
\begin{align}
\label{eq:naive_loss}
\mathcal{L}_j=-\frac{1}{|\mathcal{D}_j|}\sum_{i=1}^{|\mathcal{D}_j|}(y_i\log\hat{y}_{u,i}+(1-y_i)\log(1-\hat{y}_{u,i}))) 
\end{align}
% $$ \theta=\arg\mathop{min}\limits_{\Delta\boldsymbol{\Phi},\Delta\boldsymbol{W}}\mathcal{L}$$
Then, we minimize the loss by,
\begin{align}
\label{eq:naive_backward}
\Delta\Phi_j,\Delta W_j=\arg\min\limits_{\Delta\Phi_j,\Delta W_j}\mathcal{L}_j
\end{align}

\subsubsection{Lightweight Plugin Network} 
% \\
\
\newline
The above plugin network is similar to the residual module in ResNet, the difference is that we use the residual of the network parameters instead of the residual of the feature. Such a method can stably extract the personalized information of each group into the residual parameters $\Delta\Phi$ and $\Delta W$. However, in order to avoid overfitting, the learning rate needs to be set very low, which leads to a stable effect but limited improvement. So we improve the plugin network. Specifically, we 
insert plugin networks between the embedding layer and the feature extractor and between the feature extractor and the classifier, whose parameters are denoted by $\Phi^{'}$ and $W^{'}$, respectively. Specifically for the $j$-th user group, their parameters are $\Phi^{'}_{j}$ and $W^{'}_{j}$ respectively.
% $$ r_{u,i,t}=f(u,i,s_{u,t}|\theta+\Delta\theta)$$
% $$ \hat{y}_{u,i,t}=f(g(\boldsymbol{e}|P_e)+\boldsymbol{m_p}|W)$$
\begin{align}
\label{eq:light_forward}
\hat{y}_{u,i}=f(u,i,j,s^u|\theta,\Phi^{'}_{j},W^{'}_{j})
\end{align}

The loss function is same to Eq.~\ref{eq:naive_loss}.
Then, we backpropagate the loss on the lightweight plugin network and minimize the loss by,
\begin{align}
\label{eq:light_backward}
\Phi^{'}_{j},W^{'}_{j}=\arg\min\limits_{\Phi^{'}_{j},W^{'}_{j}}\mathcal{L}_j
\end{align}

At the same time, we continue to update the trunk network using Eq.~\ref{eq:gradient_aggregation} with a lower learning rate. To summarize briefly, we update $\Phi^{'}_{j}$ and $W^{'}_{j}$ immediately after the $j$-th mini-batch backpropagation of each batch. The trunk network is updated after all mini-batches of each batch have completed backpropagation.
The plugin network can fine-tune the intermediate features inside the trunk network to meet the needs of model personalization. The trunk network with continuous gradient aggregation can not only continuously improve the trunk network's ability to capture common information, but also prevent the bias of the plugin network from being too large.

The details of our proposed method is summarized in Algorithm~\ref{algo:main_algo}.

\vspace{-1.5mm}
\begin{algorithm}[!t]
\SetAlgoLined
Divide the dataset $\mathcal{D}$ into $\mathcal{D}_1\sim\mathcal{D}_n$ according to the level of users' activeness.\\
\textbf{Stage I:}{~\colorbox{gray!30}{$\rhd$~\emph{Train the Trunk Network}} \\
\quad 1) Initialize \textbf{Trunk Network}. \\
% $g_\cloud(\cdot)$, $g_\device(\cdot)$ and $g_\refresh(\cdot)$. \\
\quad 2) Feed each group of samples into Eq.~\eqref{eq:original_forward} for forward propagation. \\
\quad 3) Compute loss $\hat{y}_{u,i}$ via Eq.~\eqref{eq:loss_normal}. \\
\quad 4) Backpropagate via Eq.~\eqref{eq:backward_normal} and put the gradient into the gradient memory queue.
\\
\quad 5) Aggregate the gradients via Eq.~\eqref{eq:gradient_aggregation} and update $\theta$
\\
Output the parameters $\theta$ of the \textbf{Trunk Network}.
\\
}
\textbf{Stage II:}{~\colorbox{gray!30}{$\rhd$~\emph{Train the whole network}} \\
% \While{not converging}{
\quad 1) Initialize \textbf{Plugin Network} with small values. \\
\quad 2) Feed each group of samples into Eq.~\eqref{eq:light_forward} for forward propagation. \\
\quad 3) Compute loss $\hat{y}_{u,i}$ via Eq.\eqref{eq:naive_loss}. \\ 
\quad 4) Backward the loss in Eq.~\eqref{eq:naive_loss} and update $f_p(\cdot)$ and preserve the gradients on $f_t(\cdot)$.\\
\quad 5) Backpropagate via Eq.~\eqref{eq:light_backward}, update $\Phi^{'}$ and $W^{'}$, put the gradient into the gradient memory queue. \\
\quad 6) Aggregate the gradients via Eq.~\eqref{eq:gradient_aggregation} and update $\theta$
\\
}
% }
Output the parameters $\theta$ of the \textbf{Trunk Network} and the parameters $\Phi^{'}$ and $W^{'}$ of the \textbf{Plugin Network}.
\caption{\mbox{Two-stage training strategy}}
\label{algo:main_algo}
\end{algorithm}

\section{Experiments}

\begin{table*}[t]
    \caption{Statistics of the evaluation datasets.}
    % \vspace{-0.1in}
    \vspace{-1mm}
        \centering  % 把表居中
        \label{Dataset}
        % \begin{tabular}{p{35pt}<{\centering}|p{155pt}<{\centering}}
        \begin{tabular}{m{60 pt}<{\centering}|m{35 pt}<{\centering}m{35 pt}<{\centering}m{50 pt}<{\centering}m{50 pt}<{\centering}m{50 pt}<{\centering}m{45 pt}<{\centering}}
        % \begin{tabular}{p{35 pt}p{155 pt}}
        % \begin{tabular}{C{35 pt}C{155 pt}}
        % \begin{tabular}{>{\centering\arraybackslash}p{30        pt}>{\centering\arraybackslash}p{160 pt}}
        % \begin{tabular}{p{30 pt}<{\centering\arraybackslash}p{160     pt}<{\centering\arraybackslash}}
        % \begin{tabular}{>{\centering\PreserveBackslash}p{30        pt}>{\centering\PreserveBackslash}p{150 pt}}
        % \begin{tabular}{cc}
        \toprule
        Dataset & \#Users & \#Items & \#Interactions & \#Records/user & \#Records/item & \#Density \\
        \midrule
        Movielens & 6,040 & 3,706 & 1,000,209 & 165.57 & 269.89 & 4.468\% \\
        Amazon & 69,168 & 81,473 & 3,137,442 & 45.36 & 38.51 & 0.056\% \\
        Alipay & 262,446 & 4,545 & 27,849,672 & 106.12 & 6,127.54 & 2.334\% \\
        \bottomrule  
        \end{tabular}
    % \vspace{-0.2cm}
    
    \vspace{-1mm}
    \end{table*}

In this section, we conduct extensive experiments to show the superiority of the proposed framework. Specifically, we will answer the following questions.
\begin{enumerate}
    \item \textbf{RQ1:} How effective is the proposed method compared with the state-of-the-art competitors?
    \item \textbf{RQ2:} Does the proposed method actually improve the performance of tail users?
    \item \textbf{RQ3:} How do the gradient aggregation and plugin network affect the performance of the proposed method?
    \item \textbf{RQ4:} Is the proposed method effective for other mainstream recommendation models?
\end{enumerate}

\subsection{Experiment Settings}
    \subsubsection{Dataset}
    \
    \newline
    We conduct experiments on the following publicly accessible datasets: MovieLens\footnote{http://grouplens.org/datasets/movielens/}, Amazon\footnote{http://jmcauley.ucsd.edu/data/amazon/} and Alipay. The detailed statistics of these datasets are summarized in table \ref{Dataset}.\\ 
    % Taobao\footnote{https://tianchi.aliyun.com/dataset/dataDetail?dataId=109858}\\
        \textbf{Movielens.} MovieLens is a widely used public benchmark on movie ratings. In our experiments, we use movielens-1M which contains one million samples.\\
        \textbf{Amazon.} Amazon Review dataset \cite{McAuley_Targett_Shi_Hengel_2015} is a widely-known recommendation benchmark. We use the Amazon-Books dataset for evaluation.  \\
        \textbf{Alipay.} We collect a larger-scale industrial dataset for online evaluation from the AliPay platform~\footnote{https://www.alipay.com/}. Applets such as mobile recharge service are treated as items. For each user, clicked applets are treated as positives and other applets exposed to the user are negatives.

    \subsubsection{Evaluation Metrics}
    \
    \newline
    In the experiments, we use the widely adopted AUC, HitRate and NDCG as the metrics to evaluate model performance. They are defined by the following equations.
    \begin{align}
    \begin{split}
        & \mathrm{AUC}=\frac{\sum_{x_0\in \mathcal{D}_T} \sum_{x_1 \in \mathcal{D}_F}\mathds{1}[f(x_1)<f(x_0)]}{|\mathcal{D}_T||\mathcal{D}_F|}, \\
        % & \text{NDCG}@K = \sum_{u\in \mathcal{U}}  \frac 1 {|\mathcal{U}|}  \frac{2^{\mathds{1}(R_{u,g_u}\leq K)}-1}{\log_2(\mathds{1}(R_{u,g_u}\leq K)+1)},\\
        & \text{HitRate}@K = \frac{1}{|\mathcal{U}|}\sum_{u\in \mathcal{U}} \mathds{1}(R_{u,g_u}\leq K), 
    \end{split}
    \end{align}
    where $\mathds{1}(\cdot)$ is the indicator function, $f$ is the model to be evaluated, $R_{u,g_u}$ is the rank predicted by the model for the ground truth item $g_u$ and user $u$, and $\mathcal{D}_T$, $\mathcal{D}_F$ is the positive and negative testing sample set respectively.

\subsubsection{Baselines}
\
\newline
To verify the applicability, the following representative sequential modeling approaches are implemented and compared with the counterparts combined with the proposed method.
    \
    \newline
        \textbf{GRU4Rec}\cite{GRU4Rec} is one of the early works that introduce recurrent neural networks to model user behavior sequences in recommendation. \\
        \textbf{DIN}\cite{DIN} introduces a target-attention mechanism for historically interacted items aggregation for click-through-rate prediction.\\
        \textbf{SASRec}\cite{SASRec} is a representative sequential modeling method based on self-attention mechanisms. It simultaneously predicts multiple next-items by masking the backward connections in the attention map. 
        % \textbf{MeLU}\cite{MeLU}
        % \textbf{MACR}\cite{causal}
\ 
\newline
To evaluate the effectiveness on tail user modeling, the following competing methods are introduced for comparison.
\
\newline
\textbf{Agr-Rand}\cite{Agr-Rand} introduced a gradient surgery strategy to solve the domain generalization problem by coordinating inter-domain gradients to update neural weights in common consistent directions to create a more robust image classifier.\\
\textbf{PCGrad}\cite{PCGrad} is a very classic gradient surgery model that mitigates the negative cosine similarity problem by projecting the gradients of one task onto the normal components of the gradients of the other task by removing the disturbing components to mitigate gradient conflicts.\\
\textbf{Grad-Transfer}\cite{gradtrans} adjusts the weight of each user during training through resample and gradient alignment, and adopts an adversarial learning method to avoid the model from using the sensitive information of user activity group in prediction to solve the long-tail problem.
% \textbf{}

\begin{table*}[htbp]
\caption{Group level performance comparison on each group data.}
    % \vspace{-0.1in}
    \vspace{-1mm}
        \centering  % 把表居中
        \label{AUC_performance}
\setlength{\tabcolsep}{1.5mm}
\begin{tabular}{c|c|cccccc}
\toprule
Dataset                    & Group & DIN    & Agr-Rand & Grad-Transfer & PCGrad & Ours   &  \\
\midrule
\multirow{5}{*}{Movielens} & Group 1   & 0.8262 & 0.8246   & 0.8443        & 0.8375 & \textbf{0.8434} &  \\
% \cline{2-7}
& Group 2   & 0.9026 & 0.8828   & 0.9040        & 0.8964 & \textbf{0.9131} &  \\
% \cline{2-7} 
& Group 3   & 0.9072 & 0.8950   & 0.9133        & 0.9097 & \textbf{0.9247} &  \\
% \cline{2-7}
& Group 4   & 0.9059 & 0.8899   & 0.9110        & 0.9033 & \textbf{0.9227} &  \\
% \cline{2-7}
& Group 5   & 0.9105 & 0.8813   & 0.9094        & 0.9021 & \textbf{0.9251} &  \\
\hline
\multirow{5}{*}{Amazon}    & Group 1   & 0.9196 & 0.9139   & 0.9196        & 0.9205 & \textbf{0.9233} &  \\
% \cline{2-7}
& Group 2   & 0.9218 & 0.9175   & 0.9211        & 0.9220 & \textbf{0.9249} &  \\
% \cline{2-7}
& Group 3   & 0.9237 & 0.9192   & 0.9246        & 0.9247 & \textbf{0.9261} &  \\
% \cline{2-7}
& Group 4   & 0.9240 & 0.9182   & 0.9240        & 0.9259 & \textbf{0.9275} &  \\
% \cline{2-7}
& Group 5   & 0.9272 & 0.9212   & 0.9270        & \textbf{0.9303} & 0.9285 &  \\
\hline
\multirow{5}{*}{Alipay}    & Group 1   & 0.7253 & 0.6881   & 0.7159        & 0.7269 & \textbf{0.7304} &  \\
% \cline{2-7}
& Group 2   & 0.7381 & 0.7045   & 0.7296        & 0.7390 & \textbf{0.7432} &  \\
% \cline{2-7}
& Group 3   & 0.7508 & 0.7169   & 0.7415        & 0.7511 & \textbf{0.7548} &  \\
% \cline{2-7}
& Group 4   & 0.7635 & 0.7297   & 0.7544        & 0.7625 & \textbf{0.7658} &  \\
% \cline{2-7}
& Group 5   & 0.8025 & 0.7615   & 0.7889        & 0.8002 & \textbf{0.8059} &  \\
\bottomrule                        
\end{tabular}
\label{tab:group}
% \vspace{-0.2cm}
\vspace{-1mm}
\end{table*}

\subsubsection{Implementation Details}
\
\newline
\vpara{Preprocessing.} On the Alipay dataset, the dates of all samples in the dataset are from 2021-5-19 to 2021-7-10. In order to simulate the real a/b testing environment, we use the date to divide the dataset. We take the data before 0:00 AM in 2021-7-1 as the training set, and vice versa as the test set.
On Movielens and Amazon datasets, we treat the labels of all user-item pairs in the dataset as 1, and the labels of user-item pairs that have not appeared as 0. We take the user's last sample as the test set. On Movielens, we use positive samples in the training set: the ratio of negative samples = 1:4 to sample negative samples. In the test set, we refer to \cite{movlelens_sample}, so we use all negative samples of a user as the test set. In Amazon's training set, we sample negative samples with the ratio of positive samples: negative samples = 1:4, and this ratio becomes 1:99 in the test set. We also filter out all users and items in Amazon with less than 15 clicks to reduce the dataset.
On Alipay and Amazon datasets, we group by the number of samples of users. On the Movielens dataset, we group by the length of the user's click sequence

\vpara{Implementation.} In terms of hardware, our models are trained on workstations equipped with NVidia Tesla V100 GPUs. For all datasets and all models, the batch size is set to 512. The loss function is optimized by the Adam optimizer with a learning rate of 0.001 for the gradient aggregation learning stage and 0.0001 for the plugin model learning stage. The training is stopped when the loss converges on the validation set.
        
\subsection{Experiments and Results}

\subsubsection{Overall Comparison (RQ1, RQ2).} \label{sec:overallcomparison}

The comparison between our model and three state-of-the-art models on the group level and the user level are shown in Table \ref{tab:group} and Table \ref{tab:user}, respectively.

\begin{table}[h]
    \caption{User level performance comparison on all data.}
    
    \vspace{-1mm}
        \centering  % 把表居中
        \label{Performance}
    
    \setlength{\tabcolsep}{1.5mm}
    \begin{tabular}{c|c|c|c}
    \toprule
    % \hline
    
    % \multirow{2}{*}{Method} & \multicolumn{3}{c|}{Movielens-1M} & \multicolumn{3}{c|}{Amazon} & \multicolumn{3}{c}{Alipay} \\
    Method & Movielens & Amazon & Alipay \\
    % \cline{2-10} & AUC & HR@k & NDCG@k & AUC & HR@k & NDCG@k & AUC & HR@k & NDCG@k   \\
    \midrule
    % \hline
    DIN      &    
    0.8909   &    0.9165    &  0.6905 
            \\   
            
    Agr-Rand    &   
    0.8728   &    0.9124    &   0.6754   
            \\
    
    Grad-Transfer  &    
    0.8958   &     0.9177    &     0.6853  
            \\
            
    PCGrad  &    
    0.8886  &    0.9177  &   0.6897    
            \\
            
    Ours    &    
    \textbf{0.9052}  &   \textbf{0.9202}    &   \textbf{0.6924}
            \\
    \bottomrule
    % \hline
    \end{tabular}
    \label{tab:user}
    % \vspace{-0.2cm}
    \end{table}

\begin{figure}[h]
    \centering
    \includegraphics[width=0.4\textwidth]{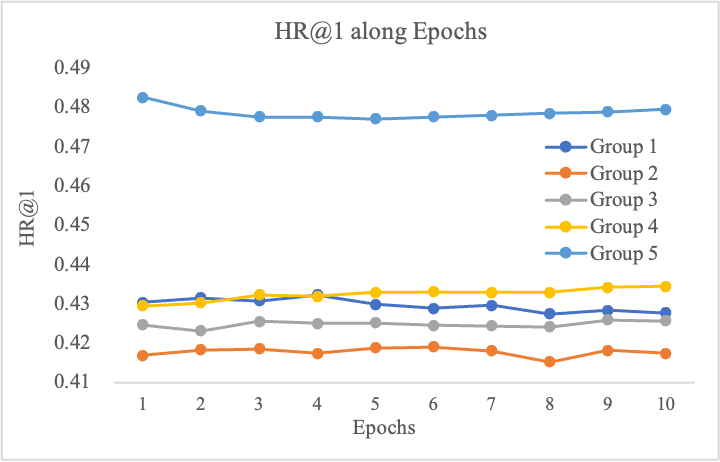}
    \vspace{-1mm}
    \caption{Performance of Hit@1 on validation set with different training epochs. The convergence points are different for different models.}
    \label{fig:hit}
    % \vspace{-0.2cm}
    
    \vspace{-1mm}
\end{figure}

% 竖表
\begin{table*}[htbp]
\caption{Performance comparison in terms of AUC to demonstrate the effects of gradient aggregation and plugin model.}
\vspace{-1mm}
\centering  % 把表居中
\label{ablation_study}
% \begin{tabular}{lllllllllll}
\setlength{\tabcolsep}{1.35mm}
\begin{tabular}{c|c|ccc|ccc|ccc}

\toprule
Dataset   & Group & GRU4Rec & +GA & +GA+PN & DIN    & +GA & +GA+PN & SASRec & +GA & +GA+PN \\
\midrule
\multirow{5}{*}{Movielens} & Group 1   & 0.8587  & 0.8595     & \textbf{0.8633}            & 0.8262 & 0.8384 & \textbf{0.8434}        & 0.8539 & 0.8591    & \textbf{0.8638}           \\
& Group 2   & \textbf{0.9264}  & 0.9210     & 0.9250            & 0.9026 & 0.8977 & \textbf{0.9131}        & 0.9242 & 0.9221    & \textbf{0.9258}           \\
& Group 3   & 0.9354  & 0.9338     & \textbf{0.9385}            & 0.9072 & 0.9023 & \textbf{0.9247}        & 0.9337 & 0.9345    & \textbf{0.9393}           \\
& Group 4   & 0.9379  & 0.9393     & \textbf{0.9424}            & 0.9059 & 0.9017 & \textbf{0.9227}        & 0.9360 & 0.9358    & \textbf{0.9399}           \\
& Group 5   & 0.9400  & 0.9371     & \textbf{0.9425}            & 0.9105 & 0.8973 & \textbf{0.9251}        & 0.9387 & 0.9355    & \textbf{0.9408}           \\
\hline
\multirow{5}{*}{Amazon}    & Group 1   & 0.9306  & 0.9338     & \textbf{0.9340}            & 0.9196 & 0.9198 & \textbf{0.9233}        & 0.9287 & 0.9329    & \textbf{0.9337}           \\
& Group 2   & 0.9329  & 0.9366     & \textbf{0.9367}            & 0.9218 & 0.9211 & \textbf{0.9249}        & 0.9315 & 0.9353    & \textbf{0.9357}           \\
& Group 3   & 0.9381  & 0.9401     & \textbf{0.9407}            & 0.9237 & 0.9216 & \textbf{0.9261}        & 0.9367 & 0.9374    & \textbf{0.9384}           \\
& Group 4   & 0.9407  & 0.9438     & \textbf{0.9452}            & 0.9240 & 0.9228 & \textbf{0.9275}        & 0.9399 & 0.9412    & \textbf{0.9417}           \\
& Group 5   & 0.9479  & \textbf{0.9487}     & 0.9470           & 0.9272 & 0.9270 & \textbf{0.9285}        & \textbf{0.9458} & 0.9444    & 0.9421           \\
\hline
\multirow{5}{*}{Alipay}    & Group 1   & 0.7312  & 0.7318     & \textbf{0.7372}            & 0.7253 & 0.7270 & \textbf{0.7304}        & 0.7283 & 0.7313    & \textbf{0.7353}           \\
& Group 2   & 0.7432  & 0.7432     & \textbf{0.7485}            & 0.7381 & 0.7397 & \textbf{0.7432}        & 0.7411 & 0.7428    & \textbf{0.7467}           \\
& Group 3   & 0.7569  & 0.7565     & \textbf{0.7620}            & 0.7508 & 0.7507 & \textbf{0.7548}        & 0.7547 & 0.7558    & \textbf{0.7609}          \\
& Group 4   & 0.7700  & 0.7694     & \textbf{0.7745}            & 0.7635 & 0.7622 & \textbf{0.7658}        & 0.7674 & 0.7682    & \textbf{0.7737}           \\
& Group 5   & 0.8103  & 0.8081     & \textbf{0.8135}            & 0.8025 & 0.7970 & \textbf{0.8059}     & 0.8053 & 0.8067    & \textbf{0.8125}           \\
\bottomrule
\end{tabular}
\label{tab:ablation}
\vspace{-1mm}
\end{table*}

\vpara{Group-level Analysis.} We are interested in how different user groups with different activeness benefit from the propose technique. Towards this end, we explicitly present the performance of these groups across three datasets. Accordingly to the results in Table \ref{tab:group}, we have the following observations:
\begin{itemize}[leftmargin=*]
    \item Overall, our technique yields the best performance across different datasets in most cases. Remarkably, the proposed method outperforms the best-performing state-of-the-art by +.0091, +.0029, and +.0042 \textit{w.r.t.} AUC on the MovieLens, Amazon, and AliPay datasets, respectively. Note that the improvement +.0042 AUC is significant for the large-scale industrial dataset, Alipay. These results demonstrate the merits of our technique for recommendation.
    \item Our method achieves consistent improvement across different user groups. Although many state-of-the-art methods achieve performance gains on some user groups, they might fail to improve the base model in some other groups. For example, Grad-Transfer achieves significant performance gains on Group1 (+.0181, AUC), but achieves inferior results on Group5 (-.0011, AUC). These results demonstrate that some state-of-the-art baselines might be at risk of hurting the experiences of some user groups. These results basically indicate the merit of our plugin components, which focus on learning personalized patterns and prevent over-correction for some users.
    \item The performance gain on tail user groups are larger than that of the head user groups in our method. Note that the major argument of this paper is that joint training in current recommendation models might be at risk of hurting tail user groups. These results reveal that our method could offer long-user groups better recommendation quality that they deserve, and importantly, without sacrificing the recommendation quality of the others. These are notable merits in industrial environments where the non-active users might be potential loyal users and the active users currently contribute the most revenue to the platform.
\end{itemize}

\vpara{User-level Analysis.} Note that the AUC in the above analysis is computed in the group level. To further reveal the performance gain in the user level, we compute the AUC metric for each user and take the average as the final performance. The results are shown in Table \ref{tab:user}. We observe that the proposed technique consistently outperforms the base model DIN and three state-of-the-art methods. It is noteworthy that although the some baselines can achieve performance gains over DIN in the group-level, they cannot beat DIN in the user level in most cases. Remember that tail users are the majority in many recommender systems. These results basically indicate that these baselines are less optimal to deal with long-tail problems, and might fail to improve the average recommendation quality of all users. On the contrary, the consistent improvement of our technique further reveals the rationality and effectiveness of our problem analysis and model design for tail user behavior modeling.

\subsubsection{Ablation Study (RQ3, RQ4).} We are interested in whether different building blocks all contribute to the proposed method. To have a comprehensive analysis, we progressively add the Gradient Aggregation component (GA) and the Plugin Network (PN) onto three base models (GRU4Rec, DIN, and SASRec) and test the constructed architectures on three datasets. The results are shown in Table \ref{tab:ablation}, which also reveal the performance change on different user groups since it is one of the primary focuses in this paper. According to the results, we observe that:
\begin{itemize}[leftmargin=*]
    \item  The proposed technique (+GA+PN) demonstrates consistent performance improvement over different base recommendation architectures across different user groups and different datasets in most cases. These comprehensive results basically indicate that the effectiveness of the proposed method is model-agnostic and dataset-agnostic. These results again verify the merits of the proposed method in long-tail sequential behavior modeling with similar findings on five user groups to those illustrated in Section \ref{sec:overallcomparison}.
    \item Not suprisingly, the Gradient Aggregation (GA) module mainly improves the recommendation performance for tail user groups while achieving comparable or sometimes inferior results compared to the base model. For example, +GA achieves AUC +.0122 on Group1 and AUC -.0132 on Group5 compared to the DIN base model on the MovieLens dataset. These results are reasonable in the sense that GA might neglect the personalization by mainly focusing on learning common information shared by all users. Nevertheless, learning common information boosts the performance on tail users, which is our primary focus, and lays the foundation for further group-specific personalization. These results basically reveal the rationality of our analysis and the effectiveness of the GA module.
    \item Based on the common information learned by the GA module, the PN module construsts an additional prediction network per group for group-specific personalization. PN achieves consistent performance improvement over different base models for five groups. These results demonstrate the necessity of personalization and the effectiveness of our design. Compared to GA, the PN componenet mostly improves the head users than tail users. For example, +GA+PN outperforms the +GA by AUC +.0278 and AUC +.005 on Group5 and Group1, respectively. These are reasonable results since tail users have significantly less number of interactions than head users. By jointly analyzing these results and those of GA, we can find that both GA and PN components are essential for bringing higher-quality recommendation to most users.
    \item According to Fig.\ref{fig:hit}, the larger the group number, the more active the user is, that is, the first group is the least active user group, and the fifth group is the most active user group. Training the plugin network for relatively inactive groups of users requires only a small number of epochs to be optimal (such as groups 1 and 2). The training curve of the user group with relatively high activity level has a more stable upward trend with the increase of epoch (such as group 3, group 4 and group 5). This is mainly due to the difference in the amount of personalized information in user groups with different levels of activity. For a user group with a large amount of data, more personalized information is required, and more epochs are needed to learn the personalized information. Otherwise, only a few epochs are required.
\end{itemize}

\section{Conclusion}
In this paper, we propose a gradient aggregation strategy. We group each data set according to activity. During training, the gradients generated by each group of data are aggregated to encourage the integration of knowledge from the perspective of optimization, and greatly improve the trunk model's ability to extract common information. In addition, we design the plugin network to guarantee the extraction of personalized information in each user group. Through the architecture of the backbone network and the plugin network, the model has both strong general reasoning ability and good individual reasoning ability. Experiments on real-world datasets demonstrate the effectiveness of the proposed method by comparing with state-of-the-art baselines.

\clearpage

\bibliographystyle{ACM-Reference-Format}
\bibliography{sample-base}

\end{document}